\documentclass[review]{elsarticle}

\usepackage{epstopdf}
\usepackage{subfigure}

\journal{Journal of \LaTeX\ Templates}

\begin{document}

\begin{frontmatter}

\title{The mechanism of individual time cost heterogeneity promotes cooperation in snowdrift game\tnoteref{mytitlenote}}

\author{Hancheng Wang\fnref{myfootnote}}

\author{Zhuozhuo Gou\fnref{myfootnote}}

\author{Yansong Deng*,\fnref{myfootnote}}

\author{Qinzhen Huang\fnref{myfootnote}}

\cortext[mysecondaryaddress]{Corresponding author. E-mail address: dengyansong@gmail.com}

\tnotetext[mytitlenote]{This work was partially supported by National Science Foundation (62073270, 61673016, 61703353), innovation Research Team of
the Education Department of Sichuan province (15TD0050), and the Graduate Innovative Research Project of
Southwest Minzu University, Project number: CX2020SZ90.}

\fntext[myfootnote]{Key Laboratory of Electronic Information of State Ethnic Affairs Commission, College of Electrical and Information Engineering, Southwest Minzu University, Chengdu, Sichuan, 610041, China.}

\begin{abstract}

Cost of time passing plays an important role when investigate the collective behaviour in real world. Each rational individual can get a more reasonable strategy by comprehensively considering the time cost. Motivated by the fact, we here propose a mechanism with individual time cost heterogeneity whose core lies in two aspects: 1. The individuals in the rule network are divided into 2 groups: high-time cost and low-time cost. 2. Each individual is endowed with a time cost parameter, and the individuals take into account the effect of time cost on the benefit when they interact with a neighbour. The synchronous updating algorithm is used to study the evolution of cooperation with time cost on a regular lattice. Simulation results show that the proposed mechanism effectively promotes cooperation in the snowdrift game. Moreover, it is revealed that the following reasons lead to a higher level of cooperation: The higher time cost, the more individuals of high-time cost in the group, and the more differences of time cost between groups when low-time cost remains unchanged.
\end{abstract}


\end{frontmatter}

\section{Introduction：}

Cooperation is the general trend in human society and biome. Research on cooperative behavior has become a hot spot in academic research \cite{ref1,ref2,ref3}. And for studying this issue, evolutionary game theory provides a useful theoretical framework which can resolve the contradiction between cooperation and selfless \cite{ref4}. The evolutionary game was developed by mathematician Von Neumann and economist Morgan Stern in 1944, and has formed many classic frameworks now. For example, snowdrift game, prisoner's dilemma game, public goods game, deer hunting game, etc \cite{ref5,ref6,ref7,ref8,ref9,ref10}.

Early in 2006, Novak proposed that cooperative behaviors should be supported by corresponding mechanisms before they can appear, and introduced five mechanisms \cite{ref11} to promote cooperation: Kinship selection, direct reciprocity, indirect reciprocity, network reciprocity and group selection. The core idea of kinship selection is to explain the cooperative behavior that occurs between individuals with kinship or closer relationships, the derived mechanism is: Hanmilotons rule \cite{ref12} (the condition for altruistic behavior is $r>c/b$, where $r$ is the probability of having the same gene, and $c/b$ is the cost-benefit-ratio. For example, $r=1/2$ between brothers, cousins between $r=1/8$), the herd mechanism (individuals will choose the strategy that their neighbors choose the most to change themselves). Direct reciprocity \cite{ref13} is the interaction that the two sides of the game will repeat many times. If this side chooses cooperation this round, then the probability of the other party choosing cooperation will be very high for the next round. There are mechanisms derived from this idea: Tit-For-Tat (TFT) rule \cite{ref14,ref15}, which completely imitates the other party's previous selection. Win-Stay, Lost-Shift (WSLS) rules \cite{ref16,ref17}. In this rule, if this round of selection gains high returns, the next round will keep the strategy, whereas the subsequent round will make the opposite choice. Indirect reciprocity \cite{ref18} arises from the complex relationship of human society. The interaction between individuals and individual views is limited, and more interaction with strangers, so it can also be explained that indirect reciprocity is based on direct reciprocity. Mechanisms derived from indirect reciprocity, such as, reward and punishment mechanism \cite{ref19,ref20,ref21,ref22,ref23}. Its basic idea is to reward cooperator and punish defectors. Both rewards and punishments can promote cooperation, but the research of various scholars and studies tells us that punishments can promote cooperation more than rewards. In the real world, an individual will not interact with all individuals, but only with a few individuals, and such an interactive relationship constitutes a network structure \cite{ref24,ref25,ref26}. Therefore, many scholars have also focused on the network structure: The first evolutionary game proposed by Novak on the regular network \cite{ref27}, and then the research on the evolutionary game on the small-world network and the scale-free network has also become a hot spot for scholars \cite{ref28,ref29,ref30,ref31,ref32,ref33,ref34,ref35,ref36}. In addition to these 5 types of mechanisms, some scholars have found that adding memory mechanism \cite{ref37,ref38,ref39} can also promote cooperation. The idea of the memory mechanism is that the individual can refer to the strategies of the previous rounds and choose the strategy in the round with the highest benefit for him to play this round of the game. Some scholars have proposed adding some parameters, such as the heterogeneity of learning ability \cite{ref40}.

In 2019, some scholars suggested that time cost in the snowdrift game can also promote the level of cooperation \cite{ref41}. They assume that two drivers are stuck on the road. If one chooses to cooperate and the other chooses to defect, the defector doesn't seen to lose anything, but the passing of time can also cause losses to the defector. But the article has certain limitations. We find that few studies focus on the influence of group’s type of the evolution of cooperation in network games. From this perspective, we propose a method for the effect of individual time cost heterogeneity(ITCH) mechanism on the level of cooperation in the snowdrift game. We mainly study the snowdrift game model, in addition to the originally considered snow shoveling cost, a new time cost parameter $t_{cost time}$ is added. Secondly, the whole group is divided into $A$(high-time cost) and $B$(low-time cost), that is to say, the time cost considered by $A$ and $B$ is different. Because, in human society, the importance of time for people with different identities is different, so the time cost for diverse groups of people due to the passing of time is also different. Considering these, we introduce a parameter $t_{cost time}$ that is the value of time cost which is represents the time cost of each individual caused by the passage of time. And introduce a parameter $V$ to represent the percentage of $A$ group in the whole group. In order to better reflect the impact of differences between groups on the level of cooperation, we also introduces a parameter $\alpha$ for research. Simulation results show that the proposed mechanism effectively promotes cooperation in the snowdrift game.

The rest of this paper is organized as follows: In Section 2, we introduce a model of ITCH mechanism. In Section 3, we show the simulation results and analyze it. In Section 4, we give the conclusion of this article.

\begin{figure}[!htb]
  \centering
  \includegraphics[width=12cm,height=9cm]{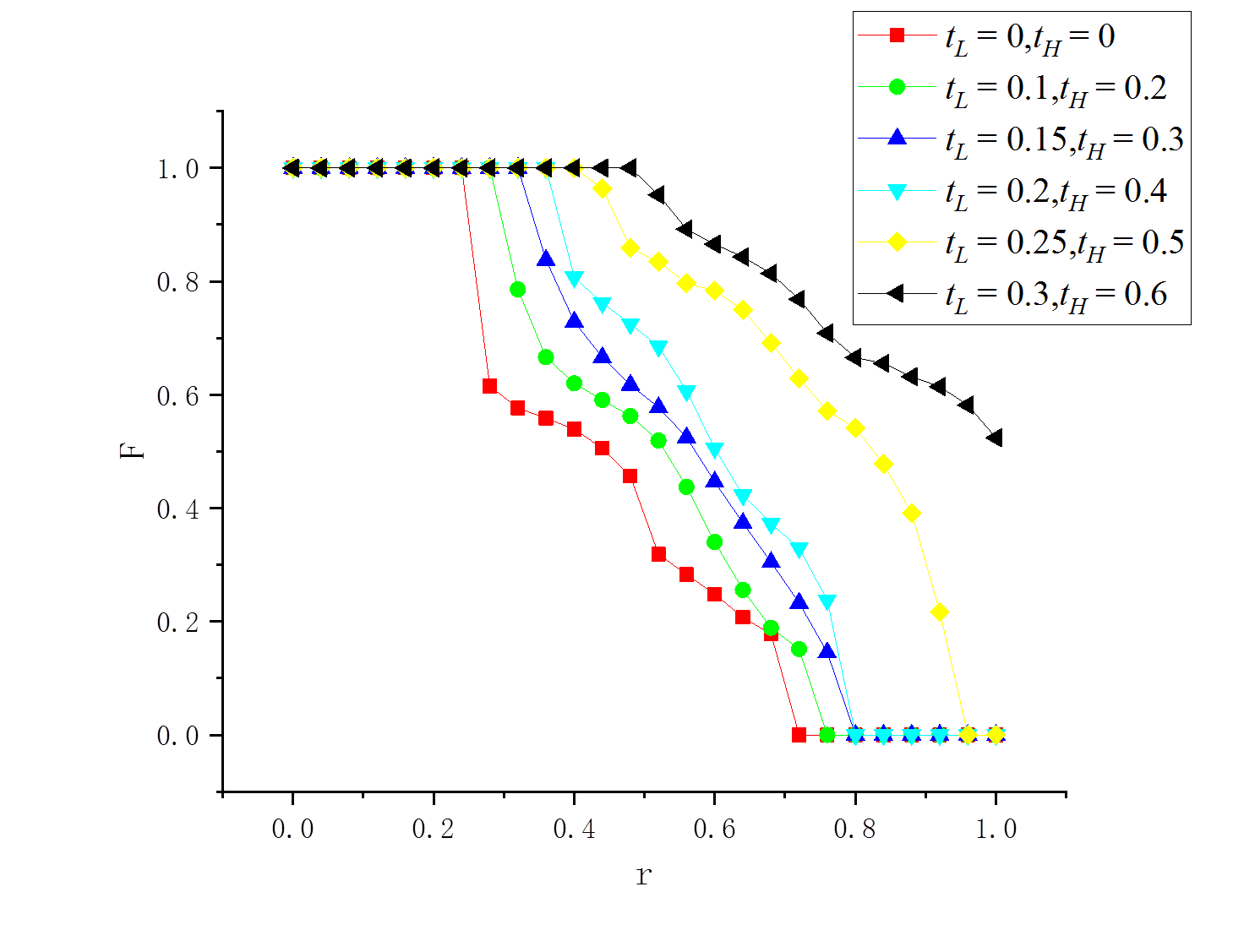}
  \caption{For the case of $V$=0.5, $\alpha$=2, with different $t_{cost time}$, the graph showing the fraction of cooperators F with the change of cost-to-benefit ratio of snow shoveling r, where $t_{L}=0, 0.1, 0.15, 0.2, 0.25, 0.3,t_{H}=0, 0.2, 0.3, 0.4, 0.5, 0.6$. When $t_{L}=0,t_{H}=0$, it is the traditional snowdrift game. The larger the $t_{cost time}$ value, the higher the level of cooperation.}
\end{figure}

\begin{figure}[!htb]
  \centering
  \includegraphics[width=12cm,height=9cm]{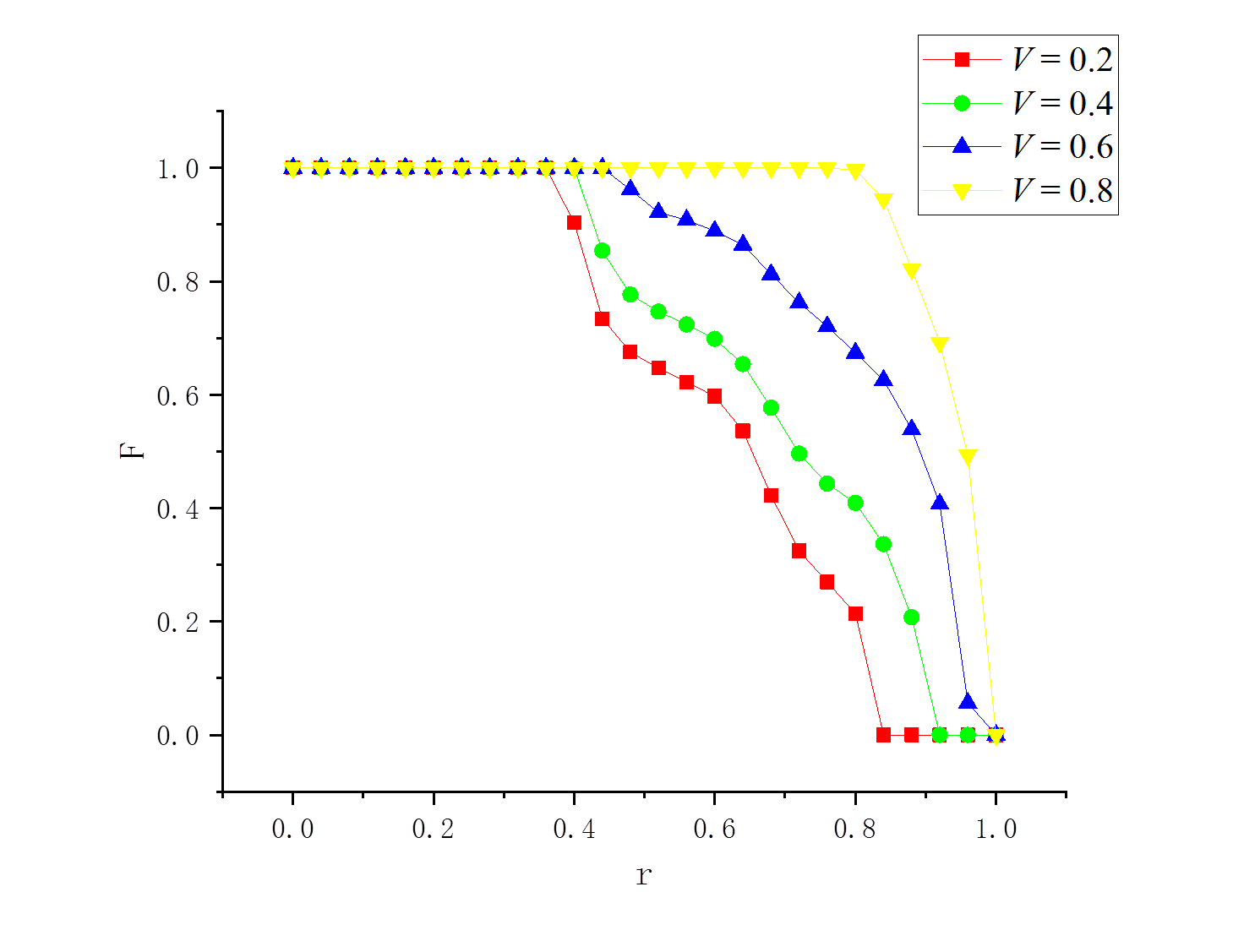}
  \caption{For the case of $t_{L}=0.25,t_{H}=0.5$, $\alpha$=2, with different $V$, the graph showing the fraction of cooperators F with the change of cost-to-benefit ratio of snow shoveling r, where $V=0.2, 0.4, 0.6, 0.8$. The larger the $V$ value, the higher the level of cooperation.}
\end{figure}

\begin{figure}[!htb]
  \centering
  \includegraphics[width=12cm,height=9cm]{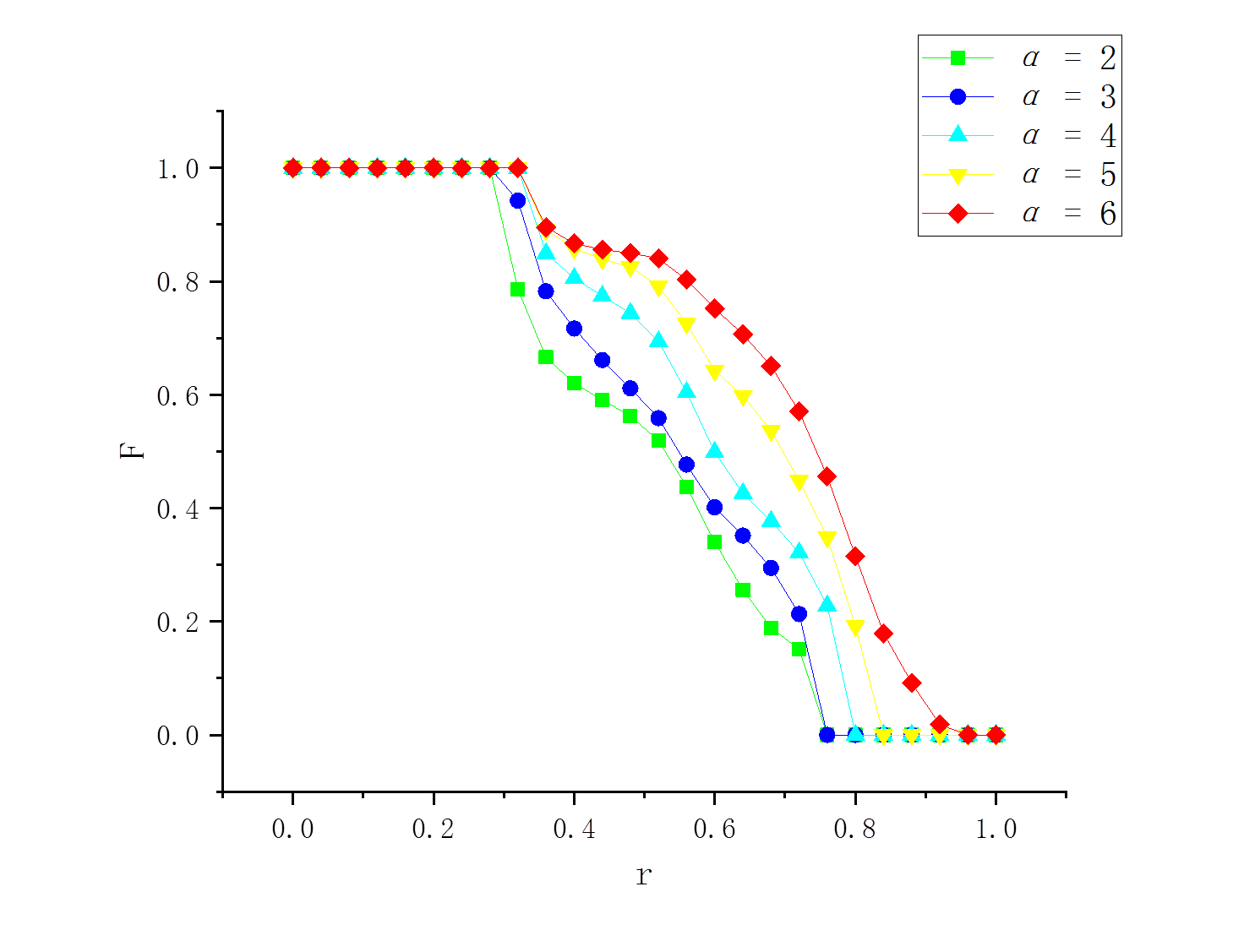}
  \caption{For the case of $t_{L}=0.1,t_{H}=0.2$, $V$=0.5, with different $V$, the graph showing the fraction of cooperators F with the change of cost-to-benefit ratio of snow shoveling r, where $\alpha=2, 3, 4, 5, 6$. The larger the $\alpha$ value, the higher the level of cooperation.}
\end{figure}

\begin{figure}[!htb]
\centering
\subfigure[$r$=0.2.]{
\includegraphics[width=4.5cm]{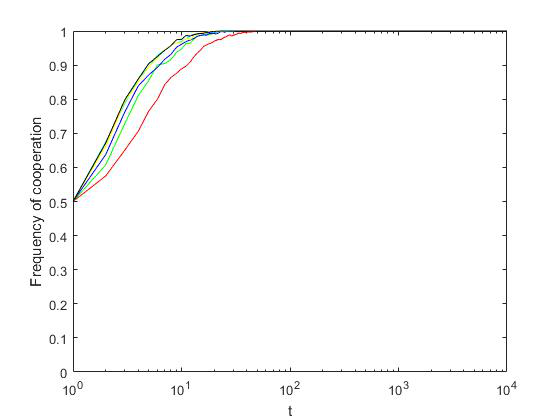}
}
\quad
\subfigure[$r$=0.4.]{
\includegraphics[width=4.5cm]{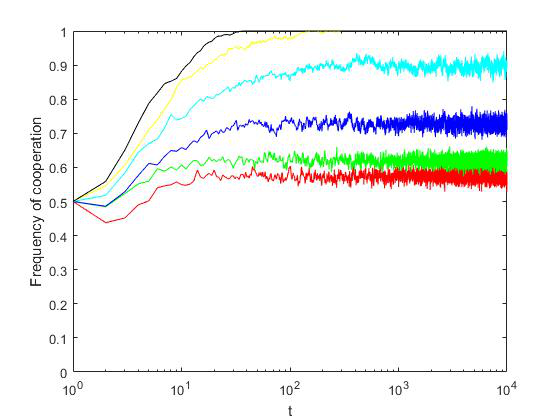}
}
\quad
\subfigure{
\includegraphics[width=1.5cm]{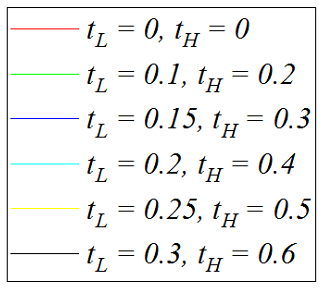}
}
\quad
\subfigure[$r$=0.6.]{
\includegraphics[width=4.5cm]{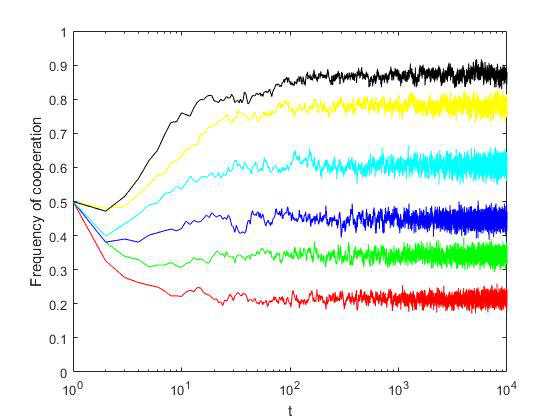}
}
\quad
\subfigure[$r$=0.8.]{
\includegraphics[width=4.5cm]{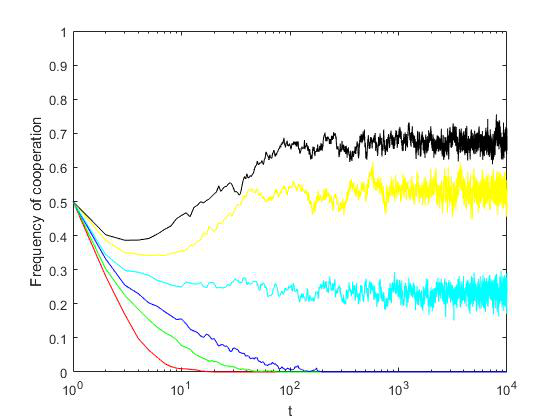}
}
\subfigure{
\includegraphics[width=1.5cm]{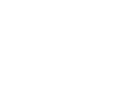}
}
\quad
\caption{When $r$=0.8, $\alpha$=2 and $V$=0.5, time evolution of the proportion of cooperators F on square lattices for different $t_{cost time}$.}
\end{figure}

\section{Model}

The individuals in the network are divided into 2 groups, $A$ (high-time cost) and $B$ (low-time cost). Individual $A$ and individual $B$ production is defined by the parameters $V$ and $(1-V)$, where $V$ represents the percentage of individual $A$ in the total population. In the traditional snowdrift game, the calculation of revenue is:
\begin{equation}       
  \begin{array}{ccc}   
     & C & D \\  
    C & b-\frac{c}{2} & b-c\\
    D & b & 0 \\   
  \end{array}
\end{equation}

Among them, $T> R> S> P$, and $2R =T+S$. According to the parameterization method, we transform the matrix into another more suitable form.
\begin{equation}       
  \begin{array}{ccc}   
     & C & D \\  
    C & 1 & 1-r\\
    D & 1+r & 0 \\   
  \end{array}
\end{equation}

Among them. $T=1+r$, $R=1$, $P=0$, $S=1-r$, and $r=\frac{c}{(2b-c)}=\frac{c}{2}$, $r\in(0,1)$ represents the cost benefit ratio of mutual cooperation.

When one side chooses to cooperate and the other side chooses to defect, the defector seen to have no cost, but as time goes by, the defector will also incur time cost. So define a parameter $t_{cost time}$ to represent time cost. In a situation where one chooses to cooperate and the other chooses to defect, the defector’s benefit will become:
\begin{center}
\begin{equation}
W=1+r-2rt_{cost time}
\end{equation}
\end{center}
In this formula, the defector's benefit minus the time cost, and $t_{cost time}$ represents the proportion of time cost in the cost $c=2r$. Different individuals have different sensitivities to the passage of time, so the benefit of the category $A$ group in the above situation is:
\begin{center}
\begin{equation}
W_{1}=1+r-2rt_{H}
\end{equation}
\end{center}
The benefit of the category $B$ group in the above situation is:
\begin{center}
\begin{equation}
W_{2}=1+r-2rt_{L}
\end{equation}
\end{center}
And $t_{H}=\alpha t_{L}$($\alpha$ is the differences of time cost coefficient between high-time cost and low-time cost)

Each point on the grid represents an individual $x$, and the probability of each individual choosing cooperation or defection strategy in the first round is 50$\%$. 4 dots around $x$ represents its 4 neighbors $y$, $y\in\{1,2,3,4\}$. In each round of the game, the individual $x$ chooses the appropriate strategy $C$ or $D$ to play with its neighbors, and at the same time it's neighbors also choose the appropriate strategy $C$ or $D$. In the model, this game is repeated four times among individuals who are not repeated among four neighbors. Then the four benefits obtained by individual $x$ are accumulated. In summary, the cumulative benefit obtained by individual $x$ in round $t$ is as follows:
\begin{center}
\begin{equation}
U_{t,x}=\sum U_{t,x,y}
\end{equation}
\end{center}

For better simulation, we use the Fermi update formula to update the strategy of individual $x$, the formula is as follows:
\begin{center}
\begin{equation}
W(S_{x}\leftarrow S_{y})=\frac{1}{1+exp[-(U_{t,x}-U_{t,y})/ K]}
\end{equation}
\end{center}
In the formula, $t$ indicates round parameter, $K (0 <K <\infty)$ is under the update rule of the noise parameter, the individual $x$ maintains its original strategy with a probability of $1-W$. The parameter $K\rightarrow\infty$, indicates that all individuals are not affected by the neighbor's benefit, and the decision is completely taken in a random way. In this case, even if the benefit of the individual $x$ is greater than the neighbor $y$, the individual will still change the strategy, which will make the benefit lower. When $K\rightarrow0$, it means that the individual is completely rational. As long as $U_{t,x}>U_{t,y}$ is satisfied, the individual $x$'s strategy is replaced by the strategy of its neighbor $y$. Depending on the above description, we choose $K=0.1$. The simulation starts in a random state and iterates continuously until the simulation results are stable.

\begin{figure}[!htb]
\centering
\subfigure[$r$=0.2.]{
\includegraphics[width=4.5cm]{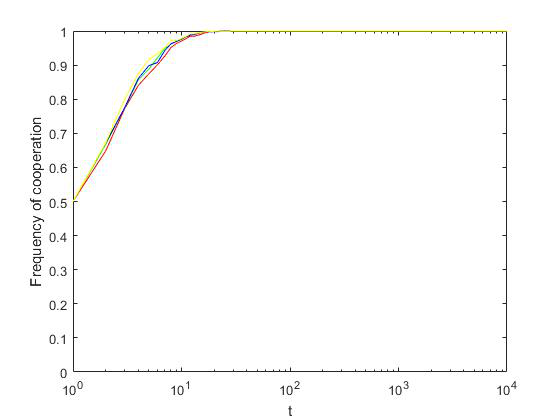}
}
\quad
\subfigure[$r$=0.4.]{
\includegraphics[width=4.5cm]{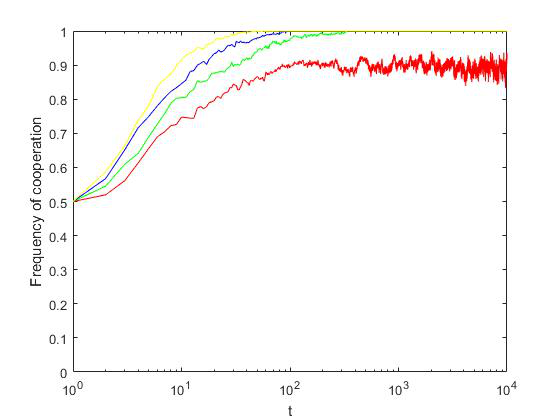}
}
\quad
\subfigure{
\includegraphics[width=1.5cm]{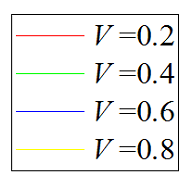}
}
\quad
\subfigure[$r$=0.6.]{
\includegraphics[width=4.5cm]{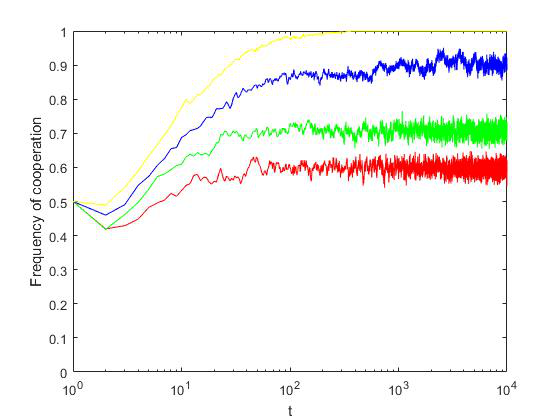}
}
\quad
\subfigure[$r$=0.8.]{
\includegraphics[width=4.5cm]{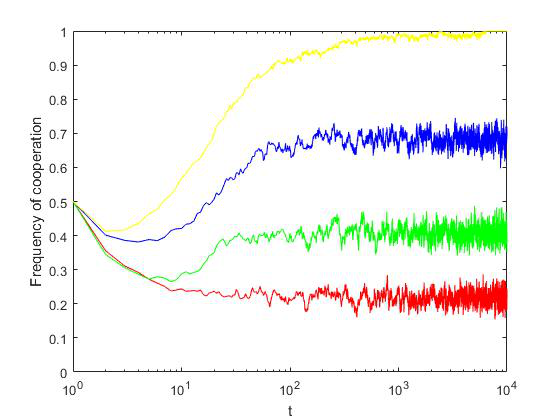}
}
\subfigure{
\includegraphics[width=1.5cm]{46.png}
}
\quad
\caption{When $r$=0.8, $\alpha$=2 and $t_{L}=0.25,t_{H}=0.5$, time evolution of the proportion of cooperators F on square lattices for different $V$.}
\end{figure}

\section{Simulation and discussion}

The simulation process is performed on a regular grid with $L \times L$ vertices. Each player is connected to his four nearest neighbors. The initial cooperative probability of each individual is set as $0.5$, the size of the regular grid is $N = L \times L = 100 \times 100 = 10000$. The average frequency of cooperative acts is quantified as the cooperation level F, where F is obtained by counting the ratio of the number of cooperators in the whole population after the system reaches a relatively steady state, that is, the F is gained by averaging $10^{3}$ rounds after a transient of $10^{4}$ rounds. In order to assure suitable accuracy, each point in simulation figures is averaged 20 interdependent realizations.

\begin{figure}[!htb]
\centering
\subfigure[$r$=0.2.]{
\includegraphics[width=4.5cm]{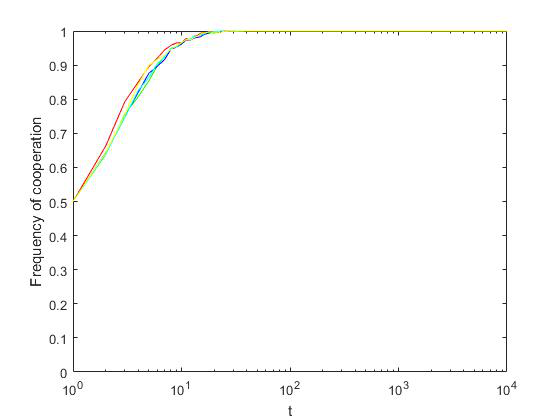}
}
\quad
\subfigure[$r$=0.4.]{
\includegraphics[width=4.5cm]{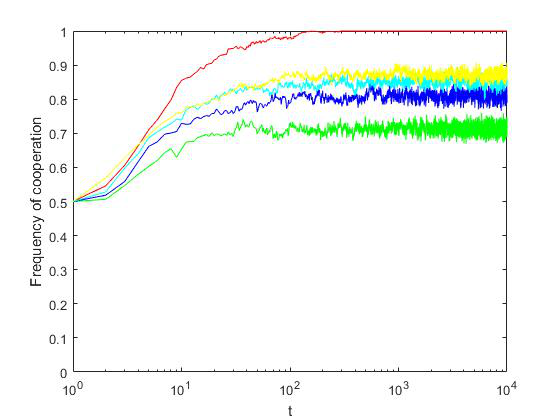}
}
\quad
\subfigure{
\includegraphics[width=1.5cm]{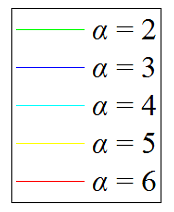}
}
\quad
\subfigure[$r$=0.6.]{
\includegraphics[width=4.5cm]{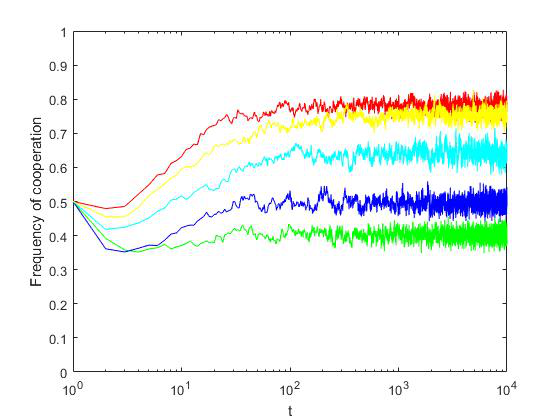}
}
\quad
\subfigure[$r$=0.8.]{
\includegraphics[width=4.5cm]{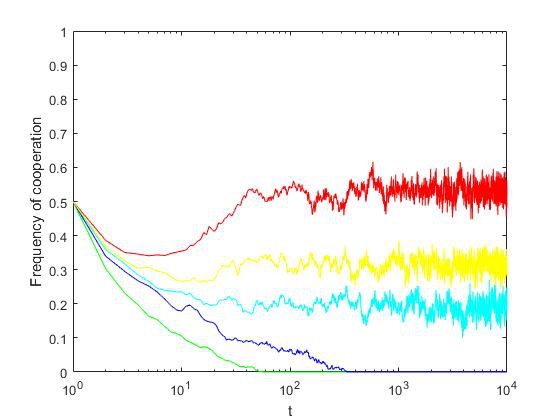}
}
\subfigure{
\includegraphics[width=1.5cm]{46.png}
}
\quad
\caption{When $r$=0.8, $V$=0.5 and $t_{L}=0.1,t_{H}=0.2$, time evolution of the proportion of cooperators F on square lattices for different $\alpha$.}
\end{figure}

In our simulation, the following fact is observed: As shown in Fig. 1,  we can see that the mixed strategy cannot maintain a high level of cooperation in the case of $t_{L}=0,t_{H}=0$ (that is, the traditional Fermi learning process), the system quickly turns into complete defect in the process of increasing $r$. However, in the case of $t_{L}\neq0,t_{H}\neq0$ (that is, considering the learning process with ITCH mechanism), the number of defectors can be greatly reduced and the level of cooperation can be significantly improved. For different time cost $t_{cost time}$, the cooperation level of the system increases as the increasing of $t_{cost time}$ value. When $t_{L}=0.1,t_{H}=0.2$, players have few cost time, then the impact on the cooperation level is small. However, compared with $t_{L}=0,t_{H}=0$, the threshold value of $r$ increases a lot, and when $r$ is about $0.78$, the system is completely betrayed. Moreover, the larger $t_{L}$ and $t_{H}$ can enhance cooperation further.

Then, we also pay attention to the situation that the cooperators' fraction F changes with cost-to-benefit ratio of snow shoveling $r$ at several different $V$ (as shown in Fig. 2). The value of F increases when $V$ continue to increase. This is because $V$ controls the benefit of group, the higher the $V$ value is, the more time cost is and the more it will take for player to pay for their cost, the more their will choose cooperate. At the same time scale, players who continue to adopt the defect strategy will gain very few payoffs, only players who continue to adopt cooperative strategies can consistently maintain high payoffs, which successfully reduces large-scale defect and improves the level of cooperation. Therefore, the higher the $V$ value, the higher the level of cooperation.

We have also studied the effect of the $\alpha$ value on the cooperation level at different payoff. $\alpha$ represents the time cost difference between the low-time cost individuals and the high-time cost individuals in the model when low-time cost unchanged. As shown in Fig. 3 that for a small $r$($0<r<0.3$), F remains at a high level. Then $r$ gradually increases, the smaller $\alpha$ is, the faster the cooperator will disappear.

\begin{figure}[!htb]
\centering
\subfigure[$t_{L}=0,t_{H}=0,t=1$.]{
\includegraphics[width=3.5cm]{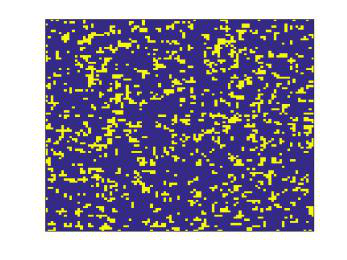}
}
\quad
\subfigure[$t_{L}=0,t_{H}=0,t=10$.]{
\includegraphics[width=3.5cm]{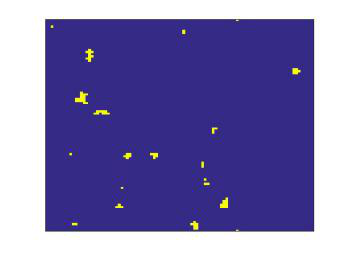}
}
\quad
\subfigure[$t_{L}=0,t_{H}=0,t=50$.]{
\includegraphics[width=3.5cm]{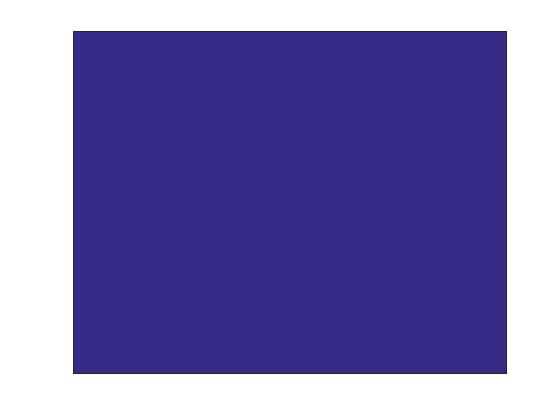}
}
\quad
\subfigure[$t_{L}=0.15,t_{H}=0.3,t=1$.]{
\includegraphics[width=3.5cm]{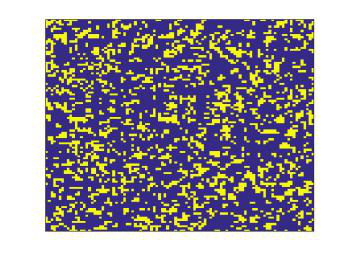}
}
\quad
\subfigure[$t_{L}=0.15,t_{H}=0.3,t=10$.]{
\includegraphics[width=3.5cm]{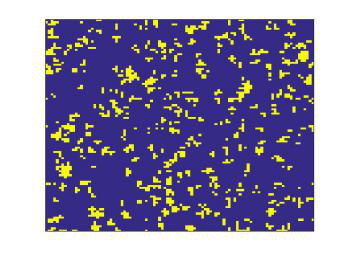}
}
\quad
\subfigure[$t_{L}=0.15,t_{H}=0.3,t=500$.]{
\includegraphics[width=3.5cm]{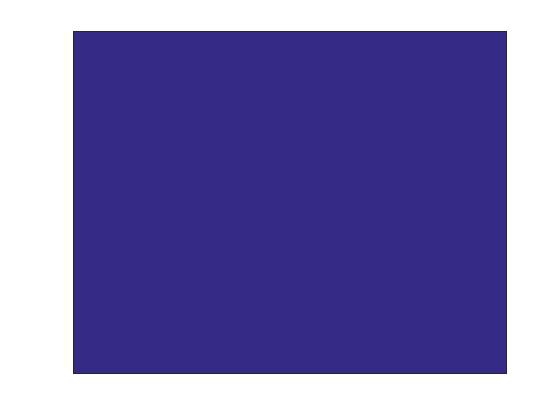}
}
\quad
\subfigure[$t_{L}=0.3,t_{H}=0.6,t=1$.]{
\includegraphics[width=3.5cm]{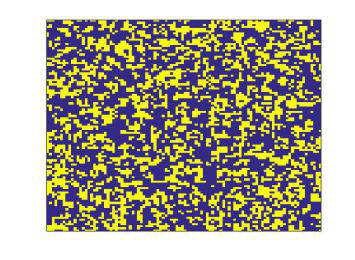}
}
\quad
\subfigure[$t_{L}=0.3,t_{H}=0.6,t=10$.]{
\includegraphics[width=3.5cm]{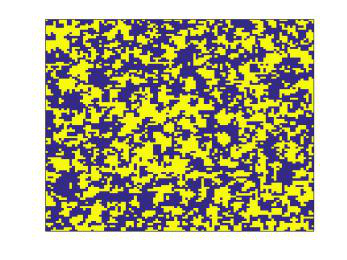}
}
\quad
\subfigure[$t_{L}=0.3,t_{H}=0.6,t=500$.]{
\includegraphics[width=3.5cm]{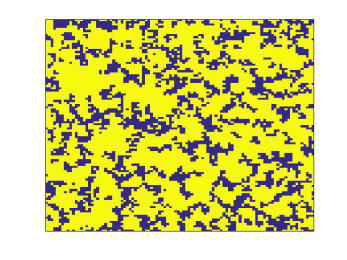}
}
\caption{A snapshot diagram of the characteristics of cooperators (yellow) and defectors (blue) at different $t_{cost time}$ and time steps. The time cost from top to bottom are $t_{L}=0,t_{H}=0$, $t_{L}=0.15,t_{H}=0.3$ and $t_{L}=0.3,t_{H}=0.6$. The results are obtained by setting $r$= 0.8, $\alpha$=2 and $V$= 0.5.}
\end{figure}

In order to further verify the obtained results, the numerical simulations of time courses that describe the relation between F and time steps are shown in Figs. 4-6.

It is found from Fig. 4 that for a small $r$=0.2, 0.4, F increases sharply. After a transitory period, F arrives at a relatively stable state. Obviously, the F for $t_{L}=0.1, 0.15, 0.2, 0.25, 0.3 ,t_{H}=0.2, 0.3, 0.4, 0.5, 0.6$ is higher than that for $t_{L}=0,t_{H}=0$. For an intermediate $r$=0.6, the F for tL,tH=0, 0.1, 0.15 decreases dramatically, while that for tL,tH = 0.2, 0.25, 0.3 surges. When the r is larger(r=0.8), the F values in the considered model still keep an upward trend with a fast speed for $t_{L}=0.3,t_{H}=0.6$ and a slight speed for $t_{L}=0.25,t_{H}=0.5$.

Then, we observe the effects of cooperation in population proportion $V$. For the $r$=0.2, 0.4, 0.6, F increases significantly for all $V$. When further increasing $r$ ($r$=0.8), the F value for $V$=0.6, 0.8 continually remains a high ratio, while that for $V$=0.2, 0.4, shows a downward trend.

Finally, let us notice the impacts of evolution of cooperation in $\alpha$ value. For the $r$=0.2, 0.4, F increases significantly for all $\alpha$. However, for a large $r$=0.8, the F for $\alpha$=2, 3, 4, 5 deceases quickly, but $\alpha$=6 decrease slowly, then gradually rise to a level.

\begin{figure}[!htb]
\centering
\subfigure[$V=0.2,t=1$.]{
\includegraphics[width=3.5cm]{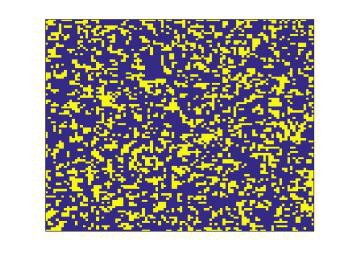}
}
\quad
\subfigure[$V=0.2,t=10$.]{
\includegraphics[width=3.5cm]{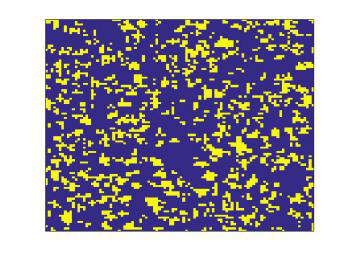}
}
\quad
\subfigure[$V=0.2,t=100$.]{
\includegraphics[width=3.5cm]{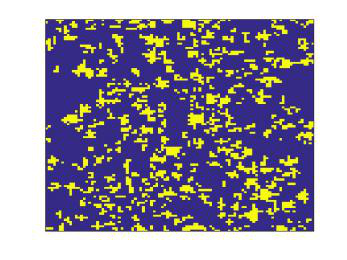}
}
\quad
\subfigure[$V=0.6,t=1$.]{
\includegraphics[width=3.5cm]{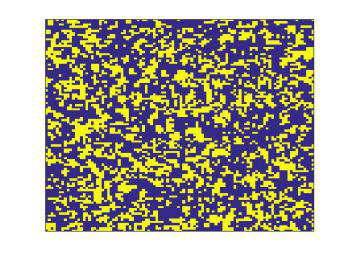}
}
\quad
\subfigure[$V=0.6,t=10$.]{
\includegraphics[width=3.5cm]{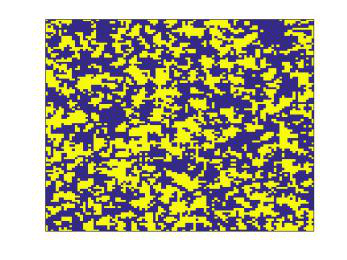}
}
\quad
\subfigure[$V=0.6,t=100$.]{
\includegraphics[width=3.5cm]{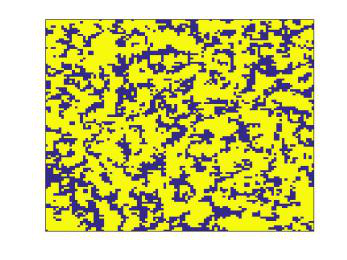}
}
\quad
\subfigure[$V=0.8,t=1$.]{
\includegraphics[width=3.5cm]{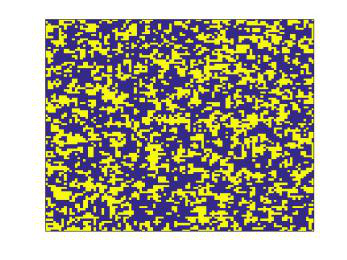}
}
\quad
\subfigure[$V=0.8,t=10$.]{
\includegraphics[width=3.5cm]{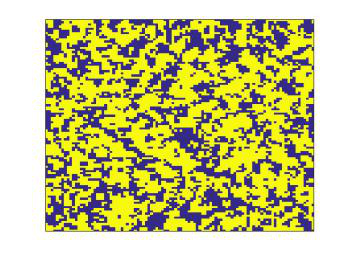}
}
\quad
\subfigure[$V=0.8,t=10000$.]{
\includegraphics[width=3.5cm]{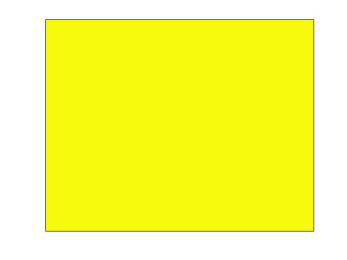}
}
\caption{A snapshot diagram of the characteristics of cooperators (yellow) and defectors (blue) at different $V$ and time steps. The time cost from top to bottom are $V=0.2$, $V=0.6 $and $V=0.8$. The results are obtained by setting $r$=0.8, $\alpha$=2 and $t_{L}=0.25,t_{H}=0.5$.}
\end{figure}

From the above descriptions, it is seen that in the three mentioned situations, whether the cost-to-benefit ratio of snow shoveling $r$ is small, intermediate or large, the F can be enhanced by the ITCH mechanism compared to the traditional model. On the other hand, to further understand the reason of above phenomena, we next scrutinize the microscopic evolutionary process in Figs. 7-9.

One can see from Fig. 7 that we choose a large $r$=0.8, when the $t_{L}=0.3,t_{H}=0.6$, the cooperators can survive by forming the compact clusters to avoid being exploited by the defectors, which will isolate the defectors little by little from the whole population. Although some patches of defectors appear in the evolutionary process, this does not prevent the extinction of them.

Similar results also appear in Fig. 8 Especially, for the large $V$=0.8, cooperators form the clusters faster to protect themselves against invasion. As a result, the defectors vanish in the lattice.

Analogous phenomenon is also discovered in Fig. 9 Nevertheless, when r is larger, the situation becomes much harsher in a small $\alpha$. A number of players are attracted by the temptation to defect and defectors win the evolutionary race.

By analyzing the evolutionary process of cooperators and defectors on a square lattice for different $r$, we elucidate the reason why the ITCH strategy can promote cooperation in the structured population to some extent.

\begin{figure}[!htb]
\centering
\subfigure[$\alpha=2,t=1$.]{
\includegraphics[width=3.5cm]{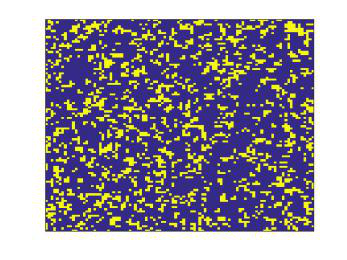}
}
\quad
\subfigure[$\alpha=2,t=50$.]{
\includegraphics[width=3.5cm]{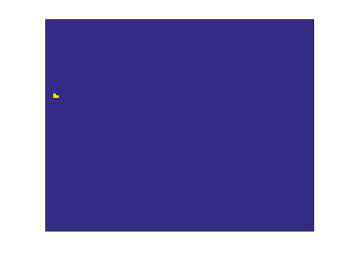}
}
\quad
\subfigure[$\alpha=2,t=100$.]{
\includegraphics[width=3.5cm]{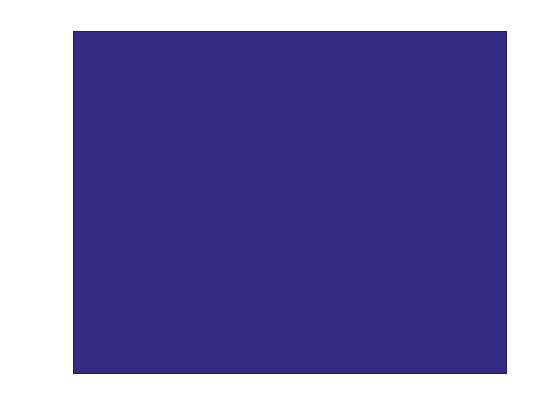}
}
\quad
\subfigure[$\alpha=4,t=1$.]{
\includegraphics[width=3.5cm]{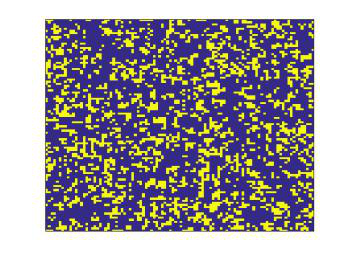}
}
\quad
\subfigure[$\alpha=4,t=50$.]{
\includegraphics[width=3.5cm]{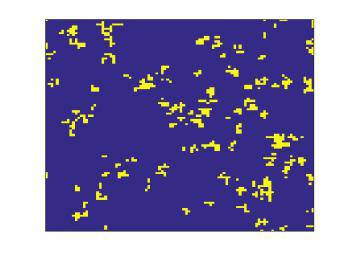}
}
\quad
\subfigure[$\alpha=4,t=100$.]{
\includegraphics[width=3.5cm]{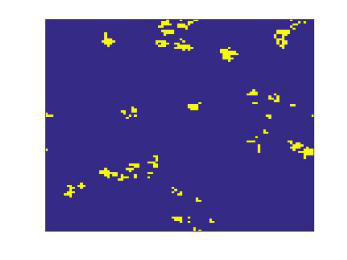}
}
\quad
\subfigure[$\alpha=6,t=1$.]{
\includegraphics[width=3.5cm]{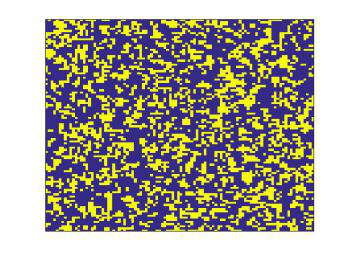}
}
\quad
\subfigure[$\alpha=6,t=50$.]{
\includegraphics[width=3.5cm]{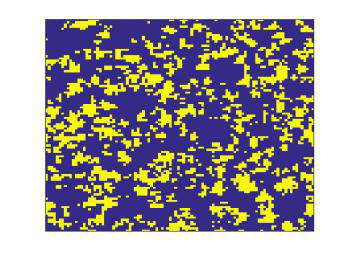}
}
\quad
\subfigure[$\alpha=6,t=100$.]{
\includegraphics[width=3.5cm]{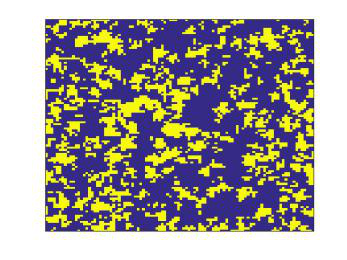}
}
\caption{A snapshot diagram of the characteristics of cooperators (yellow) and defectors (blue) at different $V$ and time steps. The time cost from top to bottom are $\alpha$=2, $\alpha$=4 and $\alpha$=6. The results are obtained by setting $r$= 0.8, $V$=0.5 and $t_{L}=0.25,t_{H}=0.5$.}
\end{figure}

\section{Conclusions}

We have intensively explored how the mechanism of ITCH affects cooperation on the regular network. And we discuss the effect of different time cost, proportions of the two groups and the time cost difference on the level of cooperation in the model.

As for the evolution, each player interacts with its nearest four neighbours. The time cost plays a crucial part in the evolution of cooperation. It is found that when time cost increases gradually, the cooperation level always increases. And the level of cooperation is also affected by the proportion of the two groups, the simulation shows that the bigger proportion of high-time cost individual in the group leads to a higher level of cooperation. In addition, we discover if low-time cost remains unchanged, the greater time cost difference, the higher level of cooperation.

In conclusion, the mechanism of ITCH promotes cooperation.

\section*{References}

\bibliography{mybibfile}

\end{document}